\documentclass[aoas,nameyear,rotating,seceqn,dvips]{arximspdf}
\usepackage{dcolumn,multirow}
\usepackage{graphicx}

\doi{10.1214/10-AOAS330}
\volume{4}
\issue{3}
\pubyear{2010}
\firstpage{1139}
\lastpage{1157}

\makeatletter
\newcommand{\citeasnoun}[1]{\citet{#1}}
\newcolumntype{d}[1]{D{.}{.}{#1}}

  \let\sv@tabnotetext\tabnotetext
  \let\sv@tabnotemark@fmt\tabnotemark@fmt
   \long\def\legend#1{{\let\tabnote@indent\leavevmode\sv@tabnotetext[]{}{#1}}}

\makeatother

\begin{document}
\begin{frontmatter}

\title{Age- and
    time-varying proportional hazards models for employment discrimination}
\runtitle{Age and time}

\begin{aug}
\author[A]{\fnms{George} \snm{Woodworth}\ead[label=e1]{george-woodworth@uiowa.edu}}
\and
\author[B]{\fnms{Joseph} \snm{Kadane}\corref{}\ead[label=e2]{kadane@stat.cmu.edu}}
\runauthor{G. Woodworth and J. Kadane}
\affiliation{University of Iowa and Carnegie Mellon University}
\address[A]{Department of Statistics\\\quad and Actuarial Sciences\\
University of Iowa\\
241 Schaefer Hall\\Iowa City, Iowa 52240\\USA\\\printead{e1}} 
\address[B]{Department of Statistics\\Carnegie Mellon University\\232 Baker Hall\\
Pittsburgh, Pensylvania  15213\\USA\\\printead{e2}}
\end{aug}

\received{\smonth{4} \syear{2009}}
\revised{\smonth{1} \syear{2010}}

\begin{abstract}
We use a discrete-time proportional hazards model of time to involuntary
employment termination. This model enables us to examine both the
continuous effect of the age of an employee and whether that effect has
varied over time, generalizing earlier work [Kadane and Woodworth
\textit{J. Bus. Econom. Statist.} \textbf{22} (2004) 182--193]. We model the log hazard surface (over age and time) as a thin-plate
spline, a Bayesian smoothness-prior implementation of penalized likelihood
methods of surface-fitting [Wahba (1990) \textit{Spline Models for Observational Data.} SIAM]. The nonlinear component
of the surface has only two parameters, smoothness and anisotropy.
The first, a scale parameter, governs the overall smoothness of the
surface, and the second, anisotropy, controls the relative smoothness
over time and over age. For any fixed value of the anisotropy parameter,
the prior is equivalent to a Gaussian process with linear drift over
the time--age plane with easily computed eigenvectors and eigenvalues
that depend only on the configuration of data in the time--age plane
and the anisotropy parameter. This model has application to legal
cases in which a company is charged with disproportionately disadvantaging
older workers when deciding whom to terminate. We illustrate the
application of the modeling approach using data from an actual
discrimination case.
\end{abstract}

\begin{keyword}
\kwd{Age discrimination}
\kwd{thin plate spline}
\kwd{smoothness prior}
\kwd{discrete proportional hazards}
\kwd{semiparametric Bayesian logistic regression}.
\end{keyword}

\end{frontmatter}

\section{Introduction}

Federal law prohibits discrimination in employment decisions on the
basis of age. There are two different bases on which a case may be
brought alleging age discrimination. First, in a disparate impact
case, the intent of the defendant is not at issue, but only the effect
of the defendant's actions on the protected class, namely, those forty
or older. For example, a rule requiring new hires to have attained
bachelor's degrees after 1995 would be facially neutral, but would
have the effect of preventing the hiring of older applicants. For
such a case, data analysis is essential to see whether the data support
disproportionate disadvantage to persons over 40 years of age with
respect to whatever employment practices might be in question. Those
practices might include hiring, salary, promotion and/or involuntary
termination. A disparate treatment case, by contrast, claims intentional
discrimination on the basis of age. Malevolent action, as well as
intention, must be shown in a disparate treatment case. While statistics
can address the defendant's actions in a disparate treatment case,
usually intent is beyond what data alone can address.

This paper uses a  proportional
hazards model as the likelihood [\citeasnoun{Cox1972}]. \citeasnoun{FikLev1994} used such a model using as dependent variable
the positive part of ($\mathit{age} -40$) as an explanatory variable. \citeasnoun{KadWoo2004}
treat age as a continuous variable, but do not model the response as a
function of calender time. This paper models both age and time
continuously. This choice enables us
to examine both the effect of age of an employee on employment decisions
(our example uses involuntary terminations) and whether that effect
has varied over time. Hence, there are two continuous variables, time
and the age of the employee. In this way, the work here generalizes
our earlier work [\citeasnoun{KadWoo2004}]
that allowed continuous time, but reduced age
to a binary variable (over 40/under 40). The analysis presented
here allows us to address the extent to which a pattern or practice
of age-based discrimination extends over a period of time. Proportional
hazards regression is particularly suited to a pattern or practice
case because it concerns the probability or odds of a person of a
given age being involuntarily terminated relative to that of a person
of another age (or range of ages), and hence directly addresses whether
an older person is disproportionately disadvantaged.

We choose to use Bayesian inference because we find that it directly
gives the probability that a person of a given age at a particular
time is more likely to be fired than another person of a given other
age at the same time. This contrasts with sampling-theory methods
that give probabilities in the sample space, even after the sample
is observed [\citeasnoun{Kadane1990a}]. When combined with sensitivity analysis,
Bayesian analysis permits us to assess the relative influence of the
data and the model. We undertook the line of research in
\citeasnoun{KadWoo2004} and in this paper to deal with
temporally-sparse employment actions taken over a long time period. We
particularly wanted to avoid the need to aggregate data into arbitrary
time periods---months, quarters, years, etc.---in order to apply
Cochran--Armitage type tests and the like.

\section{Proportional hazards regression}

The data required to analyze age discrimination in involuntary terminations
comprise the beginning and ending dates of each employee's period(s)
of employment, that employee's birth date, and the reason advanced
by the employer for separation from employment (if it occurred). Table
\ref{tab:Flow-Data} is a fragment of the data
analyzed in Section \ref{CaseWRevisited} below. Data were obtained for all persons
employed by a firm at any time between 06/07/1989 and 11/21/1993.
The tenure of the last employee shown is right censored; that is, that
employee was still in the work force as of 12/31/1993, and we are
consequently unable to determine the time or cause of his or her eventual
separation from the firm (involuntary termination, death, retirement,
etc.).
\begin{table}\tablewidth=10 cm
\caption{Flow data for the period June 1, 1989
to December 31, 1993}\label{tab:Flow-Data}
\begin{tabular*}{10 cm}{@{\extracolsep{\fill}}cccc@{}}
\hline
\textbf{Birth date}  & \textbf{Entry date}  & \textbf{Separation date}  & \textbf{Reason}\\
\hline
$\vdots$  & $\vdots$  & $\vdots$  & $\vdots$\\
3/1/1925  & 3/1/1961  & 6/1/1990  & Vol\tabnoteref{S1}\\
4/9/1938  & 4/8/1961  & 8/17/1992  & Vol\\
10/17/1934  & 4/5/1962  & 6/3/1992  & Invol\\
12/9/1939  & 4/7/1962  & 12/18/1991  & Invol\\
11/29/1932  & 5/29/1962  & 8/26/1989  & Invol\\
9/5/1928  & 10/27/1962  & 6/12/1991  & Vol\\
5/31/1941  & 1/12/1963  & n/a  & n/a \\
$\vdots$  & $\vdots$  & $\vdots$  & $\vdots$\\
\hline
\end{tabular*}
\tabnotetext[a]{S1}{``Voluntary'' termination includes death and
retirement.}
\end{table}

\subsection{Overview}

\begin{figure}

\includegraphics{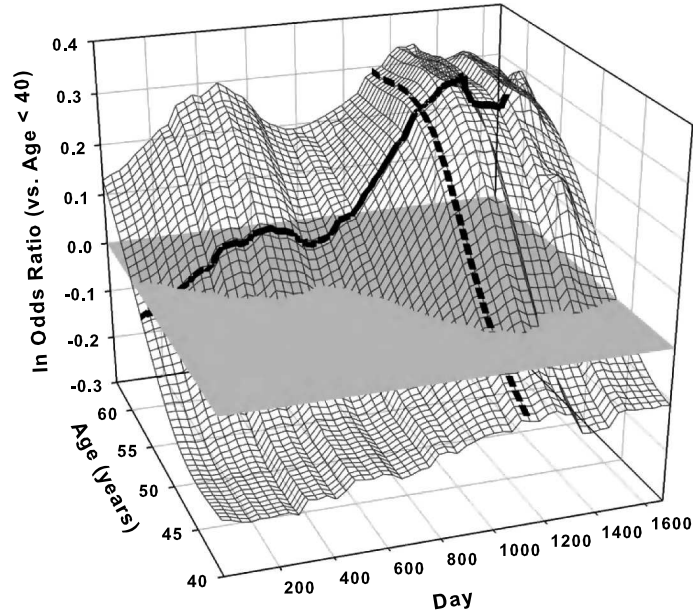}

\caption{Smooth-model-derived log odds of termination
relative to under-40 employees.}\label{fig:smooth lor}
\end{figure}

The purpose of our statistical analysis is to determine how an employee's
risk of termination depends on his or her age and how the risk for
employees of a given age changes with time. The idea is to estimate
a surface such as the one in Figure \ref{fig:smooth lor} in such
a way that it balances a penalty for infidelity to the data and for
a penalty for a surface that is unrealistically ``rough'' [\citeasnoun{Gersch1982}]. The
result is a surface that is generally within the margins of sampling
error but is also smooth. Smoothness, generally speaking, amounts
to not having areas of high curvature (i.e., spikes, cliffs, buttes,
sharp creases, etc.). The idea is to get a good fit to the data without
sacrificing smoothness.

The mesh surface in Figure \ref{fig:smooth lor} is derived from a
thin-plate spline model of the log odds (logit) of the probability of
  involuntary termination
at a given time and age. The vertical axis shows the posterior median
log-odds ratio of termination for employees of a given age on a given
date relative to the weighted average rate for employees aged 39 years
or younger on the same date (the legally unprotected class
often
used by statistical experts as a reference class
for claims of disparate impact\footnote{Note, however, that Mr. Justice Scalia's majority
  opinion in \textit{O'Connor v. Consolidated Coin Caterers Corp.}, 517
  U.S. 308 (1996) states that ``though the prohibition is limited to
  individuals who are at least 40 years of age, \S631(a). This
  language does not ban discrimination against employees because they
  are aged 40 or older; it bans discrimination against employees
  because of their age, but limits the protected class to those who
  are 40 or older. The fact that one person  in the protected class
  has lost out to another person in the protected class is thus
  irrelevant, so long as he has lost out \textit{because of his age.}''}). The gray plane corresponds to odds
ratios equal to 1.00, indicating no age discrimination relative to
the reference class; points above this plane exhibit discrimination.
Although the underlying thin plate spline is smooth, the log-odds ratio
surface is locally slightly rough because the observed numbers of
employees in each age bin at the time of each termination were used
as weights in computing the termination rate in the reference class.
%

The black ribbon in Figure \ref{fig:smooth lor} is the trajectory
of the log-odds ratio over time for employees aged 56--57, and the
dashed ribbon is the log-odds ratio as a function of age on day 1121
(05/30/92), the date of the involuntary termination of 57-year old plaintiff
W1 in Case W described in \citeasnoun{KadWoo2004}. The height of the surface at their
intersection (0.297) is the posterior median log odds on the involuntary
termination of 56--57 year-old employees relative to those under 40
on that date.

\begin{figure}

\includegraphics{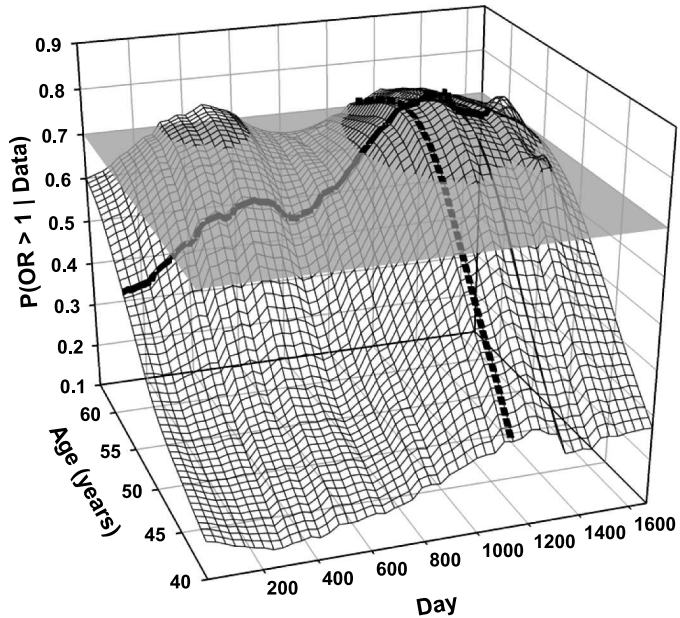}

\caption{Probability of age discrimination relative
to under-40 employees.}\label{fig:Probability-of-age}
\end{figure}

Figure \ref{fig:Probability-of-age} shows the posterior probability
of age discrimination relative to under-40 employees as a function
of age and date. Points above the gray plane represent dates and ages
at which there was at least 70\% posterior probability of age discrimination.
By itself, this would be comparatively weak evidence; however, \citeasnoun{Kadane1990b}, commenting on empirical research by \citeasnoun{MosYou1990},
suggests that this level of probability could, in standard usage,
be said to make it ``likely'' that discrimination
had occurred. The height of the surface at the intersection of the
dashed and black ribbons (0.79) is the posterior probability that
employees aged 56--57 were terminated at a higher rate compared to under-40
employees.
%

\subsection{Proportional hazards models for time to event data}

We are analyzing a group of individuals at risk for a particular type
of failure (involuntary termination) for all or part of an observation
period. The $j$th person enters the risk set at time $h_{j}$ (either
his/her date of hire or the beginning of the observation period) and
leaves the risk set at time $T_{j}$ either by failure (involuntary
termination), or for other reasons (death, voluntary resignation,
reassignment, retirement), or was still employed at the end of the observation
period. The survival function $S_{j}(t)=P(T_{j}>t)$
is the probability that the $j$th employee is still employed at time~$t$.

In practice, we rescale time and age to the unit interval $[0,1]$
and, to make computations tractable, discretize each to a finite grid;
$0=t_{0}<t_{1}<\cdots<t_{p}=1$, $0=a_{0}<a_{1}<\cdots<a_{r}=1$. Let $p_{iw}$ be the conditional probability that employee (worker)
$w$ is terminated in the interval $(t_{i-1},t_{i}]$ given the parameters
and given that s(he) was in the workforce at time $t_{j-1}$. The discretized
data for this employee are $f_{iw},\ldots,f_{pw};, r_{iw},\ldots,r_{pw},$
where $r_{iw}=1(0)$ if the employee was (not) in the work force (risk
set) at time $t_{i-1}$, and $f_{iw}=1(0)$ if the worker was (not)
involuntarily terminated (fired) in that interval. The joint likelihood
for all employees is $\prod_{w=1}^{W}\prod_{i=1}^{p}p_{iw}^{f_{iw}}(1-p_{iw})^{r_{iw}-f_{iw}}$,
where $W$ is the total number of employees. Letting $a_{w}(t)$
denote the age of employee $w$ at time $t$, we use the natural parametrization $\operatorname{logit}(p_{iw})=\beta(t_{i},a_{w}(t_{i}))$,
where $\beta(t,a)$ is a smooth function of time and
age.

The aggregated data $n_{ij}$ and $x_{ij}$ are, respectively, the
number of employees with ages in the interval $[a_{j-1},a_{j})$ at
time $t_{i}$ and the number of those who were terminated in that
interval. At this level of aggregation, the likelihood is
\begin{equation}\label{eq:L(beta)}
l(\beta)=\prod_{i=1}^{p}\prod_{j=1}^{r}\exp\bigl(\beta_{ij}x_{ij}-n_{ij}\ln{\bigl(1+\exp(\beta_{ij})\bigr)\bigr)},
\end{equation}
 where $\beta_{ij}=\beta(t_{i},a_{j})$. We assume that
the grid is fine enough and the function smooth enough that variation
of $\beta$ within a grid cell is negligible. Changing the grid
requires recomputing the cell counts, $(n_{ij},x_{ij})$ and basis
vectors defined below, which is fairly time consuming. We did a few
runs with a grid roughly twice as fine (which quadrupled the run time
and storage requirements) without observing substantive changes in the
results; however, we focused our sensitivity analysis on varying the
prior distribution of the smoothness parameter, which appeared to have
much greater impact on the results. We compute the log-odds
ratio at time $t_{i}$ for employees aged $a_{j}$ relative to unprotected
employees (i.e., employees under age 40) as
\begin{equation}\label{eq:LORv40}
\beta_{ij}-
\operatorname{logit}\biggl(\sum_{\mathit{age_{u}}\leq40}{n_{iu}p_{iu}}\bigg/\sum_{\mathit{age_{u}}\leq40}{n_{iu}}\biggr),
\end{equation}
where $\mathit{age_{u}}$ is age in years corresponding to scaled value $a_{u}$,
and $\operatorname{logit}(p_{ij})=\beta_{ij}$.

\subsection{Thin-plate spline smoothness priors}

Likelihood measures fidelity to data (the larger the better); however,
it does not incorporate our belief that the hazard ratio varies comparatively
smoothly with time and age; this is provided by a roughness penalty
(the smaller the better) that is subtracted from the log-likelihood
%
\begin{equation}\label{eq:Penalty}
\frac{\lambda}{2}\int\!\!\!\int\biggl[\biggl(\frac{\partial^{2}\beta(t,a)}{\partial^{2}t}\biggr)^{2}+
2\biggl(\frac{\partial^{2}\beta(t,a)}{\partial t\,\partial a}\biggr)^{2}+
\biggl(\frac{\partial^{2}\beta(t,a)}{\partial^{2}a}\biggr)^{2}\biggr]\,dt\,da.
\end{equation}

The smoothness parameter, $\lambda,$ weights the importance of smoothness
relative to fidelity to noisy data (larger values of the smoothness
parameter produces smoother fitted surfaces). However, there is no
reason to expect the log odds to be \textit{isotropic}---equally
smooth in time and age---and for that reason we assume that there
is a rescaling $T=t/\sqrt{1+\rho^{2}},$ and $A=\rho a/\sqrt{1+\rho^{2}},$
such that the function $b(T,A)=\beta(T\sqrt{1+\rho^{2}},A\sqrt{1+\rho^{2}}/\rho)$
is equally smooth (isotropic) in $A$ and $T$. That is, the roughness
penalty is
\begin{equation}\label{eq:AnisoPenalty}
\qquad \frac{\lambda}{2}\int\!\!\!\int\biggl[\biggl(\frac{\partial^{2}b(T,A)}{\partial^{2}T}\biggr)^{2}+2\biggl(\frac{\partial^{2}b(T,A)}
{\partial T\,\partial A}\biggr)^{2}+\biggl(\frac{\partial^{2}b(T,A)}{\partial^{2}A}\biggr)^{2}\biggr]\,dT\,dA,
\end{equation}
 which reduces to the anisotropic roughness penalty,
\begin{eqnarray}\label{eq:ReducedAnisoPenalty}
&&\frac{\tilde{\lambda}}{2}\int\!\!\!\int\biggl[\biggl(\frac{\rho^{2}}{1+\rho^{2}}\frac{\partial^{2}\beta(t,a)}{\partial^{2}t}\biggr)^{2}\nonumber\\ [-8pt]\\ [-8pt]
&&\hphantom{\frac{\tilde{\lambda}}{2}\int\!\!\!\int\biggl[
}{}+2\biggl(\frac{\rho}{1+\rho^{2}}\frac{\partial^{2}\beta(t,a)}{\partial t \partial a}\biggr)^{2}+\biggl(\frac{1}{1+\rho^{2}}\frac{\partial^{2}\beta(t,a)}
{\partial^{2}a}\biggr)^{2}\biggr]\,dt\,da,\nonumber
\end{eqnarray}
where $\rho$ is called the anisotropy parameter and $\tilde{\lambda}=\lambda\rho^{3}/(1+\rho^{2})$.
When $\rho=1$ the surface is isotropic, and as $\rho\rightarrow\infty$ (or
$\rho\rightarrow0$), there is relatively less constraint on roughness
in the age (or time) dimension.

It is interesting to compare this model to the earlier one of
\citeasnoun{FikLev1994}, which is a special case of ours. In
their case, our function $\beta(\cdot, \cdot)$ takes the form
\[
\beta(t_i, a_w (t_i))=\bigl(a_w(t_i)-40\bigr)^+.\nonumber
\]
Since that function has zero second partial derivatives (except at 40,
where they do not exist), their function imposes smoothness in our
sense. One could think of this computationally as setting $\lambda
=0$.

Since the likelihood depends on the smooth function $\beta(t,a)$
only through the values $\beta_{ij}$, the roughness penalty is minimized
for fixed $\beta_{ij}$ when $\beta(t,a)$ is the interpolating
thin-plate spline with values $\beta(t_{i},a_{j})=\beta_{ij}$.
We have from Wahba [(\citeyear{Wahba1990}), page 31, equation (2.4.9)] that there exist coefficients
$c$ such that the isotropic thin plate spline $b(T,A)$
can be represented as
\begin{equation}\label{eq:GreensRep}
b(T,A)=\sum_{ij}{c_{ij}H(T-T_{i},A-A_{j})+l(T,A)},
\end{equation}
where $l(T,A)$ is an arbitrary linear function, $H(\mathbf{v})=|\mathbf{v}|^{2}\ln(|\mathbf{v}|)/(8\pi)$,
and the coefficients $c_{ij}$ satisfy the conditions $\sum_{ij}{c_{ij}}=\sum_{ij}{t_{i}c_{ij}}=\sum_{ij}{a_{j}c_{ij}}=0$.
Then the isotropic roughness penalty, equation (\ref{eq:AnisoPenalty}),
reduces to $\lambda\mathbf{c}'\mathbf{K}_{\rho}\mathbf{c}$, where $\mathbf{c}$ is
the vector of coefficients and $\mathbf{K}_{\rho}$ is
the $pr\times pr$ symmetric matrix with elements of the form $k_{ij,uv}=H(T_{i}-T_{u},A_{j}-A_{v})=H(\frac{(t_{i}-t_{u})}{\sqrt{1+\rho^{2}}},\frac{\rho(a_{j}-a_{v})}{\sqrt{1+\rho^{2}}})$.
To accommodate the constraints on vector $\mathbf{c}$, let $\mathbf{P}$
be the projection onto the linear space orthogonal to the constraints
so that $\mathbf{c=Pc}$.

Finally, let $\mathbf{PK}_{\rho}\mathbf{P}=\mathbf{U}_{\rho}\bolds{\Lambda}_{\rho}\mathbf{U}_{\rho}'$
be the spectral decomposition of $\mathbf{P}\mathbf{K}_{\rho}\mathbf{P}$ and define
the basis vectors $\mathbf{B}_{\rho}$ as the nonzero columns of
$\mathbf{U}_{\rho}\bolds{\Lambda}_{\rho}^{\mathrm{1/2}}$. It follows that the model for the vector of logits is
\begin{eqnarray}\label{eq:betaRep1}
\bolds{\beta} & = &\mathbf{K}_{\rho}\mathbf{c}+\mathbf{L}\tilde{\bolds{\phi}}\nonumber\\
 & = &\mathbf{K}_{\rho}\mathbf{Pc}+\mathbf{L}\tilde{\bolds{\phi}}\\
 & = & \mathbf{P}\mathbf{K}_{\rho}\mathbf{Pc}+(\mathbf{I}-\mathbf{P})\mathbf{K}_{\rho}\mathbf{Pc}
 +\mathbf{L}\tilde{\bolds{\phi}},\nonumber
 \end{eqnarray}
where $\bolds{\beta}$ is the matrix with $ij$th row $\beta_{ij}$
and the $ij$th row of matrix $\mathbf{L}$ is $(1,t_{i},a_{j})$.
But $\mathbf{I}-\mathbf{P}$ is the projection onto the column space of $\mathbf{L}$
and, consequently, $(\mathbf{I}-\mathbf{P})\mathbf{K}_{\rho}\mathbf{P}\mathbf{c}$
can be absorbed into the linear term. Therefore, the model reduces to
\begin{eqnarray}\label{eq:betaModel2}
\bolds{\beta} & = &\mathbf{P}\mathbf{K}_{\rho}\mathbf{P}\mathbf{c}+(\mathbf{I}-\mathbf{P})\mathbf{K}_{\rho}\mathbf{P}\mathbf{c}+\mathbf{L}\tilde{\bolds{\phi}}\nonumber\\
 & = & \mathbf{U}_{\rho}\bolds{\Lambda}_{\rho}^{1/2}(\bolds{\Lambda}_{\rho}^{1/2}\mathbf{U}_{\rho}\mathbf{c})+\mathbf{L}\bolds{\phi}\\
 & = &\mathbf{B}_{\rho}\bolds{\delta}+\mathbf{L}\mathbf{\phi},\nonumber
 \end{eqnarray}
where $\delta=\bolds{\Lambda}_{\rho}^{1/2}\mathbf{U}_{\rho}\mathbf{c}$
and
 $\mathbf{B}_{\rho}=\mathbf{U}_{\rho}\bolds{\Lambda}_{\rho}^{1/2}$.
Thus, for a given anisotropy, $\rho$, the columns of $\mathbf{B}_{\rho}$
are basis vectors for the nonlinear part of the logit vector $\bolds{\beta}$.

The roughness penalty is $\lambda\mathbf{c}'\mathbf{K}_{\rho}\mathbf{c}=\lambda\mathbf{c}'\mathbf{P}\mathbf{K}_{\rho}\mathbf{P}\mathbf{c}=
\lambda\mathbf{c}'\mathbf{U}_{\rho}\bolds{\Lambda}_{\rho}\mathbf{U}\mathbf{c}=\lambda\bolds\delta'\bolds\delta$.
The standard Bayesian interpretation of penalized likelihood estimation
is that the penalty function is the log of the prior density of $\bolds{\delta}$.
Consequently, the components of that vector are a-priori independent
and identically distributed normal random variables with precision
$\lambda$. It follows that the prior conditional variance of $\bolds\beta$
given $\lambda$, $\rho$ and $\phi$ is
\begin{eqnarray*}
\operatorname{Var}(\mathbf{B}_{\rho}\bolds\delta) & = & \lambda^{-1}\mathbf{B}_{\rho}\mathbf{B}_{\rho}'\\
& = & \lambda^{-1}\mathbf{P}\mathbf{K}_{\rho}\mathbf{P}
\end{eqnarray*}
and, consequently, if $\mathbf{d}$ is a vector such that $\mathbf{d}'\mathbf{L}=\mathbf{0},$
then
\begin{equation}\label{eq:VarDbeta}
\operatorname{Var}(d'\beta)=\lambda^{-1}\mathbf{d}'\mathbf{K}_{\rho}\mathbf{d}.
\end{equation}

The posterior distributions of $\lambda$ and $\rho$ are not well identified
by the data and it is necessary to be somewhat careful about specifying
their priors. However, the regression coefficients, $\bolds{\phi}$,
of the linear component do not influence smoothness, are well identified
by the data, and can be given diffuse, normal prior distributions.

Viewing both time and age as continuous variables allows a more precise
and general view of a firm's policy.
However, due to the comparative sparseness of the data,
some constraint on or penalty for roughness is needed to avoid an
unrealistically rough model, unlike that depicted in Figure \ref{fig:smooth lor}. It is,
of course, possible to introduce discrete discontinuities into an
otherwise smooth model at time points where there is other evidence of
a shift in employment practices [see, e.g., Figure~6 in
\citeasnoun{KadWoo2004}]. However, we do not think that it is
appropriate to ``mine'' for unknown numbers of discontinuities at
unknown time points in the sparse data common in age-discrimination cases.
Hence, it is necessary to smooth the data. The key parameters in doing
so are smoothness and anisotropy. The smoothness parameter controls
the average smoothness of the surface and the anisotropy parameter
controls the relative degree of smoothing in the age and time coordinates.

\section{Case W revisited}\label{CaseWRevisited}

Over an observation period of about 1600 days the workforce at a firm
was reduced by about two thirds; 103 employees were involuntarily
terminated in the process. A new CEO took control at day 862, near
the middle of the observation period. The plaintiff asserted that
employees aged 50 and above were targeted for termination under the
influence of the new CEO. Here we present a fully Bayesian analysis
with smoothly time- and age-varying odds ratio. The personnel data
were aggregated by status (involuntarily terminated, other) into one-week
time intervals and two-year age intervals (20--21, 22--$23,\ldots, 64$--65). Figures \ref{fig:smooth lor} and \ref{fig:Probability-of-age} show posterior medians and posterior
probabilities of age-related discrimination (i.e., of increased odds
of termination relative to unprotected employees).

\subsection{Forming an opinion about smoothness and anisotropy}\label{sub:Forming-an-Opinion}

\begin{figure}

\includegraphics{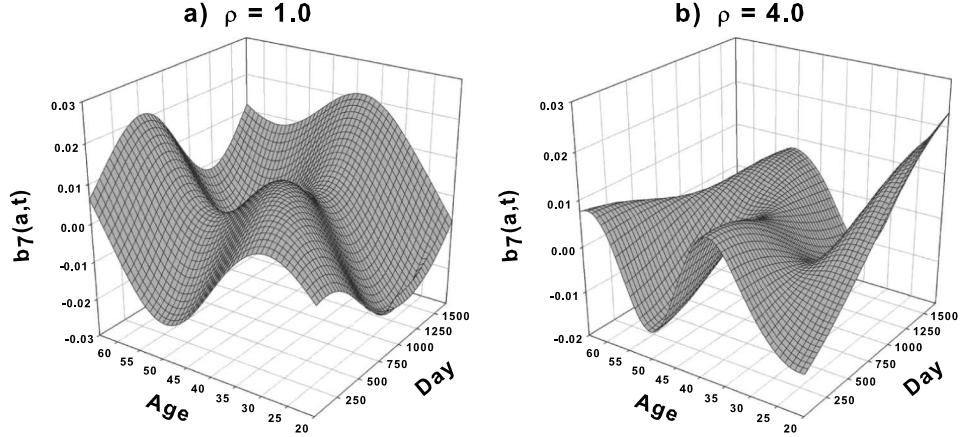}

\caption{Effect of anisotropy on the 7th basis function.}\label{fig:Effect-of-anisotropy}
\end{figure}

The anisotropy parameter $\rho$ governs the relative smoothness in
time relative to age. This is clearly illustrated in Figure \ref{fig:Effect-of-anisotropy},
which shows the seventh eigensurface (basis function) for (a) the isotropic
case where there is about one cycle in either direction in contrast
to (b) the anisotropic case $\rho=4$ in which the surface is four
times rougher in the age dimension (there are about 3 half cycles
in the age dimension to about $3/4$ of a half cycle in the time dimension).
%

In the context of employment discrimination, we think that, in terms
of roughness of the logit, a 3-year age difference is about equivalent
to a business quarter. Recalling that we have rescaled 1600 calendar
days and a 45-year age span into unit intervals, a quarter is 0.056
and a three-year age interval is 0.067 of the unit interval, corresponding
to anisotropy $\rho=1.2$. We have found empirically that doubling
or halving anisotropy has a fairly modest effect on surface shape;
consequently, we used the prior distribution shown in Table \ref{tab:Anisotropy Prior},
which has prior geometric mean 1.4.
%
\begin{table}[b]\tablewidth=6.70 cm
\caption{Prior distribution of the anisotropy parameter}\label{tab:Anisotropy Prior}
\begin{tabular*}{6.70 cm}{@{\extracolsep{\fill}}lcccccc@{}}
\hline
$\bolds\rho$ &\textbf{8}&\textbf{4} &\textbf{2} &\textbf{1} &\textbf{0.5}
&\textbf{0.25}\\
\hline
Prior & 0.08 & 0.16 & 0.26 & 0.26 & 0.16 & 0.08\\
\hline
\end{tabular*}
\legend{Larger $\rho$-values favor smoothness in time.}
\end{table}

As in our earlier analysis of this case [\citeasnoun{KadWoo2004}],
we now derive a prior
distribution for the smoothness parameter from our belief that the
odds ratio on termination for a 10-year age difference are unlikely
to change more than 15\% over a business quarter. This implies that
a particular mixed difference is unlikely to exceed 0.15 in absolute
value; that is, $\operatorname{Prior}(l|\Delta_{t}^{2}\Delta_{a}\beta(t_{0},a_{0})|\leq0.15)$
is large, where
\begin{eqnarray*}
&&\Delta_{t}^{2}\Delta_{a}\beta(t_{0},a_{0})\\
&&\qquad=\beta(t_{0}+2d_{t},a_{0}+d_{a})-2\beta(t_{0}+d_{t},a_{0}+d_{a})+\beta(t_{0},a_{0}+d_{a})\\
&&\qquad\quad{}-\beta(t_{0}+2d_{t},a_{0})+2\beta(t_{0}+d_{t},a_{0})-\beta(t_{0},a_{0}),
\end{eqnarray*}
where $d_{t}$ is a rescaled half-quarter and $d_{a}$ is a rescaled
decade. We have from equation (\ref{eq:VarDbeta}) that the
prior distribution of $\Delta_{t}^{2}\Delta_{a}\beta(t_{0},a_{0})$
is normal with mean zero and conditional variance, $\mathbf{d}'\mathbf{H}\mathbf{d}/\lambda=\mathbf{V}_{\rho}/\lambda$,
where $\mathbf{H}$ is the matrix with entries $H(T_{i}-T_{i'},A_{j}-A_{j'})$,
$\mathbf{d}$ is the vector $(1,-2,1,-1,2,-1)$, $T_{i}=(t_{0}+td_{t})/\sqrt{1+\rho^{2}}, i=0,1,2$,
and $A_{j}=\rho(a_{0}+jd_{a})/\sqrt{1+\rho^{2}},j=0,1$.
Values of $V_{\rho}$ are listed in Table \ref{tab:PriorParmsLambda}.

\begin{table}\tablewidth=253pt
\caption{Prior variance $\times\,\lambda$ of $\Delta_{t}^{2}\Delta_{a}\beta(t_{0},a_{0})$
and prior scale parameter of $\lambda$}\label{tab:PriorParmsLambda}
\begin{tabular*}{253 pt}{@{\extracolsep{\fill}}ld{1.6}d{1.2}@{}}
\hline
\multicolumn {1}{@{}l}{\textbf{Anisotropy} $\bolds\rho$}& \multicolumn {1}{c}{$\bolds{V_{\rho}}$}
& \multicolumn {1}{c@{}}{$\bolds{\mathit{sc}_{\bolds\rho}}$ \textbf{for} $\bolds{\mathit{sh}_{\bolds\rho}=0.5}$ \textbf{and} $\bolds{\alpha=0.05}$}\\
\hline
8 & 0.000383 & 5.04\\
4 & 0.000453 & 4.26\\
2 & 0.000492 & 3.93\\
1 & 0.000449 & 4.30\\
0.5 & 0.000332 & 5.81\\
0.25 & 0.000195 & 9.90\\
\hline
\end{tabular*}
\end{table}

The conditional prior distribution of the smoothness parameter given
the anisotropy parameter is gamma with shape parameter and scale parameter
selected so that $\operatorname{Prior}(|\Delta_{t}^{2}\Delta_{a}\beta(t_{0},a_{0})|\leq0.15)=1-\alpha$
is large. To complete the derivation, we have, conditional on $\rho$,
that
\[
[\Delta_{t}^{2}\Delta_{a}\beta(t_{0},a_{0})]^{2}\backsim V_{\rho}\cdot\frac{\mathit{sc}_{\rho}\Gamma(0.05)}{\Gamma(\mathit{sh}_{\rho})}\sim V_{\rho}\cdot\mathit{sc}_{\rho}\frac{1-\beta(\mathit{sh}_{\rho},0.05)}
{\beta(\mathit{sh}_{\rho},0.05)},
\]
where, abusing the notation somewhat, we let $\Gamma(\mathit{sh})$
denote an independent gamma-distributed random variable with shape
parameter $\mathit{sh}$, and let $\beta(\mathit{sh},0.5)$ denote a beta-distributed
random variable. Consequently, if
\[
\operatorname{Prior}\bigl([\Delta_{t}^{2}\Delta_{a}\beta(t_{0},a_{0})]^{2}\le0.15^{2}\bigr)=1-\alpha,
\]
then
\[
\mathit{sc}_{\rho}=\frac{0.15^{2}\beta_{\alpha}(\mathit{sh}_{\rho},0.5)}{V_{\rho}(1-\beta_{\alpha}(\mathit{sh}_{\rho},0.5))},
\]
where $\beta_{\alpha}(\mathit{sh}_{\rho},0.5)$ is the $\alpha$th
quantile of the $\beta(\mathit{sh}_{\rho},0.5)$ distribution. The
third column of Table \ref{tab:PriorParmsLambda} shows the values
of the scale parameter, $\mathit{sc}_{\rho}$ that we used to compute the surface
in Figures \ref{fig:smooth lor} and \ref{fig:Probability-of-age}.
%

\subsection{Computing the posterior distribution}

To estimate this model, we included enough basis vectors in the last
row of equation (\ref{eq:betaModel2}) to account for at least 95\%
of the total roughness variance a priori (i.e., we included basis
vectors accounting for 95\% of the sum of the eigenvalues of $K_{\rho}$).
We computed the posterior distribution of the probabilities of involuntary
termination, and of the odds ratios relative to under-40 employees
in each time--age bin using a program written in SAS IML language.
For a given anisotropy value, $\rho$, we used the Metropolis--Hastings
within the iteratively reweighted least squares algorithm proposed by
\citeasnoun{Gamerman1997} to separately update the logistic regression coefficient
vectors $\phi$ and $\delta$, and a Gibbs step to update the smoothness
parameter, $\lambda$. Anisotropy values were chosen from the six
shown in Table \ref{tab:Anisotropy Prior}, where, beginning with
an arbitrary initial value, we attempted a jump from the current anisotropy
value to an adjacent value with transition probabilities from the
6 $\times$ 6 doubly stochastic matrix shown in Table \ref{tab:Jump-proposal-probabilities}.
Letting current parameter values be $\bolds\delta$, $\bolds\phi$,
$\lambda$, and $\rho$, we attempt a reversible jump, $\rho\rightarrow\tilde{\rho}$.
We then propose values $\tilde\phi=\phi$, and $\tilde\lambda=\rho\cdot sc/\widetilde{sc}$,
where $sc$ and $\widetilde{sc}$ are scale parameters from Table
\ref{tab:PriorParmsLambda} corresponding to $\rho$ and $\tilde{\rho}$, respectively. Finally,
we generate a proposal for $\tilde{\bolds{\delta}}$ as follows.
Let $\bolds{\beta}=\mathbf{B}_{\rho}\delta+\mathbf{L}\cdot\phi$
be the current logit vector and let $\mathbf{p}$ be the current vector
of termination probabilities in time--age bins [i.e., $\operatorname{logit}(\mathbf{p})=\bolds{\beta}$]
and let $\mathbf{q}=1-\mathbf{p}$. Let vectors $\mathbf{n}$ and
$\mathbf{y}$ be the numbers at risk and terminated in the time--age
bins. Then, $\tilde{\bolds{\delta}}$ is proposed from the multivariate
normal distribution with precision $\widetilde{\bolds{\Pi}}=[\tilde{\lambda}+\mathbf{B}_{\tilde{\rho}}'\mathbf{n}\mathbf{p}\mathbf{q}\mathbf{B}_{\tilde{\rho}}]$
and mean $\tilde{\bolds{\mu}}=\widetilde{\bolds{\Pi}}^{-1}\mathbf{B}_{\tilde{\rho}}'\mathbf{n}\mathbf{p}\mathbf{q}\cdot\widehat{\mathbf{y}},$
where $\mathbf{B}_{\rho}$ is the matrix of basis vectors corresponding
to anisotropy $\rho$, as defined in the paragraph after equation
(\ref{eq:betaModel2}), and $\widehat{\mathbf{y}}=\mathbf{B}_{\rho}\delta+(\mathbf{y}-\mathbf{p})/\mathbf{p}\mathbf{q}$.
The proposal is accepted with probability
\begin{eqnarray*}
\alpha &=&\min\biggl[1,\frac{p(\tilde{\rho})p(\tilde{\lambda}|\tilde{\rho})p(\delta|\tilde{\lambda})l(\tilde{\beta})}{p(\rho)p(\lambda|\rho)p(\delta|\lambda)l(\beta)}
\cdot\frac{p(\tilde{\rho}\to\rho)q(\delta|\tilde{\lambda},\tilde{\delta},\phi)}{p(\rho\to\tilde{\rho})q(\tilde{\delta}|\lambda,\delta,\phi)}\cdot\frac{\partial\tilde{\lambda}}
{\partial\lambda}\biggr]\\
&=&\min\biggl[1,p(\tilde{\rho})\tilde{\lambda}^{\tilde{q}/2}\exp\biggl(-\frac{1}{2}\tilde{\lambda}\delta'\tilde{\delta}\biggr)l(\tilde{\beta})\\
&&\hphantom{\min\biggl[1,}\times|\Pi|^{0.5}\exp\biggl(-\frac{1}{2}(\delta-\mu^{\prime})\Pi(\delta-\mu)^{\prime}\biggr)
\\&&\hphantom{\min\biggl[1,}\Big/\biggl(p(\rho)\lambda^{q/2}\exp\biggl(-\frac{1}{2}\lambda\delta'\delta\biggr)l(\beta)
\\&&\hphantom{\min\biggl[1,\bigl(\biggl(}\times|\widetilde{\Pi}|^{0.5}\exp\biggr(-\frac{1}{2}(\widetilde{\delta}-\tilde{\mu})^{\prime}\tilde{\Pi}(\tilde{\delta}-\tilde{\mu})^{\prime}\biggr)\biggr)\biggr],
\end{eqnarray*}
where $l(\bolds\beta)$ is the likelihood function
[equation (\ref{eq:L(beta)})], $q$ and $\tilde{q}$ are the ranks of $B_{\rho}$
and $B_{\tilde{\rho}}$, and $\mu$ and $\Pi$ are the mean and precision
of the reverse proposal [\citeasnoun{Green1995}].

\subsection{Sensitivity analysis}\label{sub:sensitivity}

It is a good statistical practice to investigate whether and to what
extent the results of an analysis are sensitive to the prior distribution.
That means in this case investigating the influence of the prior distribution
of the smoothness and anisotropy parameters. Figures \ref{fig:smooth lor} and \ref{fig:Probability-of-age} above
are based on our preferred prior distribution as specified in Tables
\ref{tab:Anisotropy Prior} and \ref{tab:PriorParmsLambda}. In Figure \ref{fig:Effect-of-anisotropy} we compare Figure \ref{fig:smooth lor} (a) with an
analysis (b) in which the scale parameters in Table \ref{tab:Jump-proposal-probabilities} are multiplied
by 10, decreasing the roughness penalty by a factor of 10 and producing
a substantially rougher surface. Figure \ref{fig:Effect on Prob} shows
the effect of this variation on the probability of discrimination.
\begin{table}\tablewidth=213 pt
\caption{Jump proposal probabilities
for the anisotropy parameter}\label{tab:Jump-proposal-probabilities}
\begin{tabular*}{213 pt}{@{\extracolsep{\fill}}ld{1.1}d{1.1}d{1.1}d{1.1}d{1.1}d{1.1}@{}}
\hline
\multicolumn{1}{@{}l}{\textbf{Anisotropy}} & \multicolumn{1}{c}{\textbf{8}} & \multicolumn{1}{c}{\textbf{4}}
& \multicolumn{1}{c}{\textbf{2}} & \multicolumn{1}{c}{\textbf{1}} & \multicolumn{1}{c}{\textbf{0.5}} &
\multicolumn{1}{c@{}}{\textbf{0.25}}\\
\hline
8 & 0.9 & 0.1 &  &  &  & \\
4 & 0.1 & 0.8 & 0.1 &  &  & \\
2 &  & 0.1 & 0.8 & 0.1 &  & \\
1 &  &  & 0.1 & 0.8 & 0.1 & \\
0.5 &  &  &  & 0.1 & 0.8 & 0.1\\
0.25 &  &  &  &  & 0.1 & 0.9\\
\hline
\end{tabular*}
\end{table}
%
\begin{figure}[b]

\includegraphics{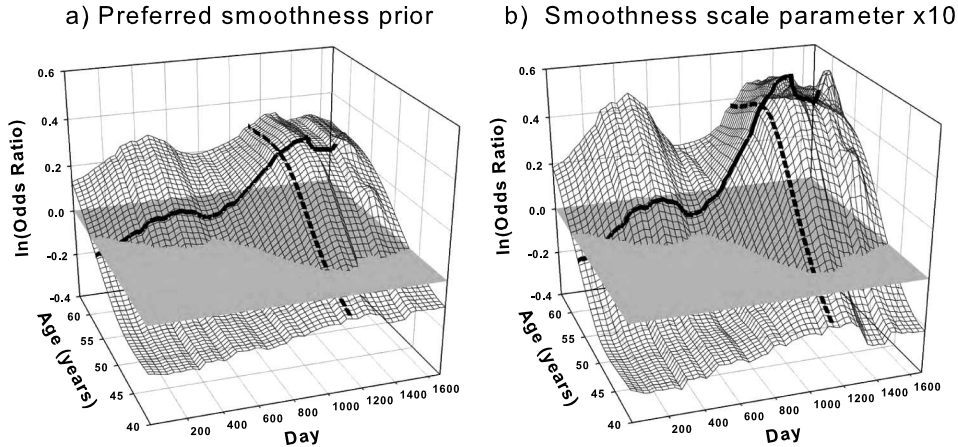}

\caption{Effect of the smoothness prior on the log
odds ratio.}\label{fig:Effect on LOR}
\end{figure}
\begin{table}\tablewidth=326 pt
\caption{Posterior distribution and marginal likelihood
of the anisotropy parameter}\label{tab:Post Dist rho}
\begin{tabular*}{326 pt}{@{\extracolsep{\fill}}d{1.2}d{1.2}d{1.3}d{1.3}d{1.2}d{1.2}d{1.2}d{1.2}d{1.2}@{}}
\hline
 &  & \multicolumn{3}{c}{\textbf{Posterior}\tabnoteref{S5}} & \multicolumn{3}{c@{}}{\textbf{Marginal
 likelihood}}\\[-7 pt]
& & \multicolumn {3}{c}{\hrulefill}&\multicolumn {3}{c@{}}{\hrulefill}\\
\multicolumn{1}{@{}l}{$\bolds\rho$} & \multicolumn{1}{c}{\textbf{Prior}} & \multicolumn{1}{c}{$\bolds{P(\rho|\mathit{Data})}$ }& \multicolumn{1}{c}{$\bolds{p_{0.025}}$}
& \multicolumn{1}{c}{$\bolds{p_{0.975}}$ }& \multicolumn{1}{c}{$\bolds{\propto P(\mathit{Data}|\rho)}$} & \multicolumn{1}{c}{$\bolds{p_{0.025}}$}&
\multicolumn{1}{c@{}}{$\bolds{p_{0.975}}$}\\
\hline
8 & 0.08 & 0.122 & 0.12 & 0.13 & 1.53 & 1.47 & 1.61\\
4 & 0.16 & 0.231 & 0.22 & 0.24 & 1.44 & 1.40 & 1.50\\
2 & 0.26 & 0.286 & 0.28 & 0.30 & 1.10 & 1.07 & 1.14\\
1 & 0.26 & 0.217 & 0.21 & 0.23 & 0.83 & 0.81 & 0.87\\
0.5 & 0.16 & 0.101 & 0.10 & 0.11 & 0.63 & 0.61 & 0.67\\
0.25 & 0.08 & 0.043 & 0.04 & 0.05 & 0.54 & 0.50 & 0.59\\
\hline
\end{tabular*}
\tabnotetext[a]{S5}{$p_{0.025}$ and $p_{0.975}$ are Monte-Carlo error bounds (see text).}
\end{table}
\begin{figure}[b]

\includegraphics{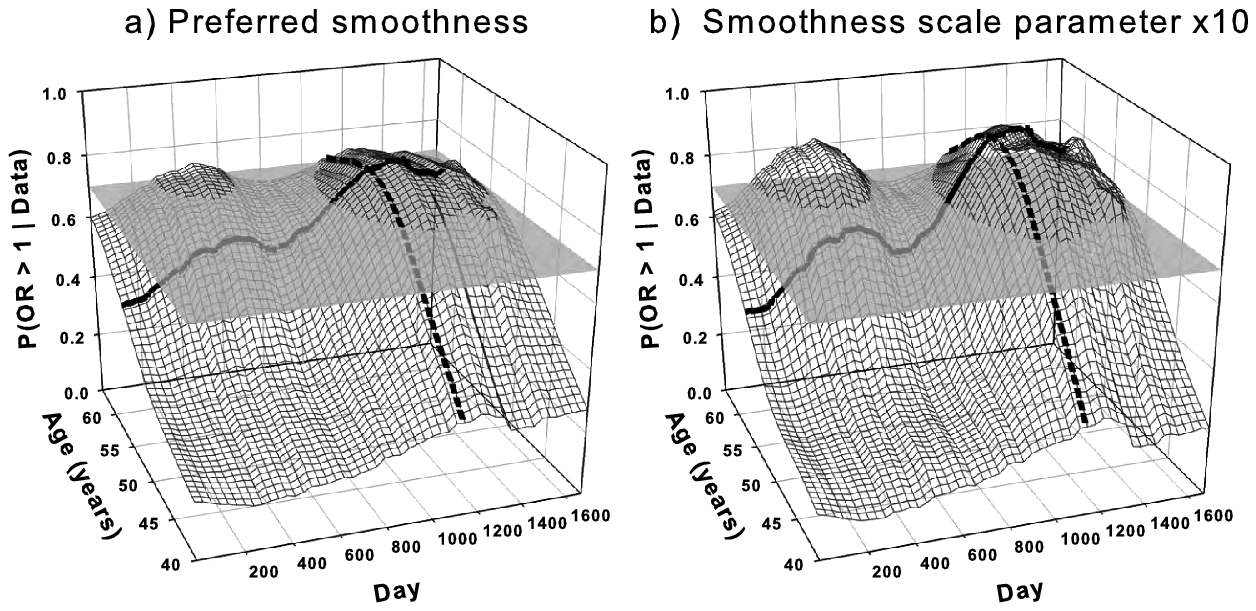}

\caption{Effect of the smoothness prior on the posterior
probability of discrimination.}\label{fig:Effect on Prob}
\end{figure}
%
\subsection{Identification of the anisotropy parameter}\label{sub:anisotropy}

Table \ref{tab:Post Dist rho} shows the\break marginal posterior
distribution of the anisotropy parameter for the preferred prior
distribution of the smoothness parameter (Table
\ref{tab:PriorParmsLambda}). The posterior probability
$P(\rho|\mathit{Data})$ is the observed rate of sampler visits to
value $\rho$ of the anisotropy parameter in 19,000 replications, the
marginal likelihood is
$P(\rho|\mathit{Data})/P(\rho)\varpropto
P(\mathit{Data}|\rho)$, and $p_{0.025}$ and $p_{0.975}$ are nominal
Monte-Carlo error bounds computed on the assumption that the observed
rate has a binomial distribution.

It is clear from the marginal likelihood that the data carry information
about anisotropy and, in particular, that models with large values
of $\rho$ (i.e., which are very rough in the time dimension) are disconfirmed
by the data. However, high levels of smoothness in the time dimension
are not disconfirmed by data and apparently must be discouraged by
the prior. Because of this, we investigated the effect of a prior
that forces more smoothness in the time dimension.

\begin{figure}

\includegraphics{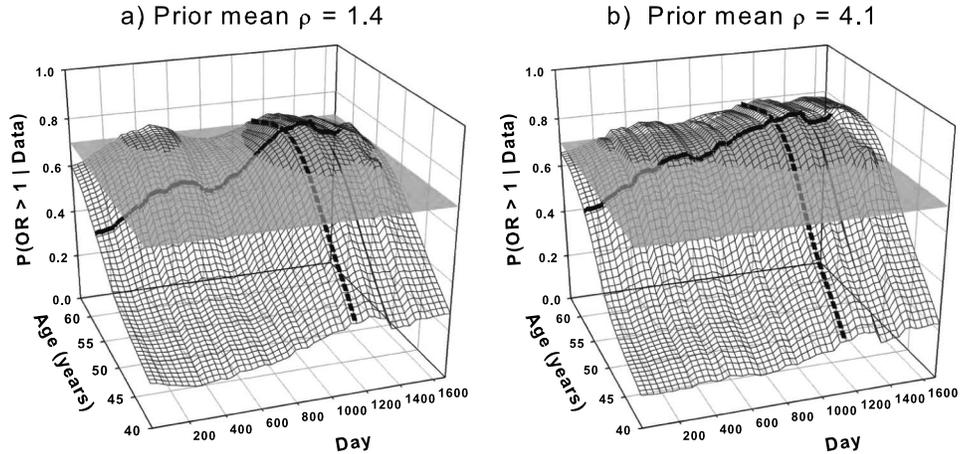}

\caption{Effect of the anisotropy parameter on
the posterior probability of discrimination.}\label{fig:Effect-of-Aniso}
\end{figure}

In Figure \ref{fig:Effect-of-Aniso} we altered the prior distribution
for the anisotropy parameter to favor smoothness in the time dimension
(Table \ref{tab:Anisotropy Prior II}). In this case the prior geometric
mean of the anisotropy parameter is about 4, meaning that we think
that, in terms of roughness of the log odds on termination, a decade
of age is about equivalent to a business quarter (see Section \ref{sub:Forming-an-Opinion}).
Evidence of discrimination in the plaintiff's case (the intersection
of the dashed and black ribbons) is slightly stronger for the prior
that forces more smoothness in the time dimension; $P$$(OR>1|\mathit{Data})$
is about 0.79 for the preferred prior (a) and about 0.83 for the more
time-smoothing prior (b).

Although the analysis in panel (b) is more favorable to the plaintiff,
we think it would be less persuasive to the trier(s) of fact (judge
or jury) since it does not seem to distinguish between the periods
before and after the arrival of the new CEO (day 862).
%
\begin{table}[b]\tablewidth=224 pt
\caption{Alternate prior distribution of the anisotropy parameter}\label{tab:Anisotropy Prior II}
\begin{tabular*}{224 pt}{@{\extracolsep{\fill}}lcccccc@{}}
\hline
$\bolds\rho$ & \textbf{8} & \textbf{4} & \textbf{2} & \textbf{1} & \textbf{0.5} & \textbf{0.25}\\
\hline
Prior & 0.5 & 0.25 & 0.125 & 0.0625 & 0.03125 & 0.03125\\
\hline
\end{tabular*}
\legend{Larger $\rho$-values favor smoothness in time.}
\end{table}
%

\subsection{Previous analyses of case W}\label{sub:previous}

The plaintiff who was between 50 and 59 years of age was one of 12
employees involuntarily terminated on day 1092. He brought an age
discrimination suit against the employer under the theory that the
new CEO had a pattern of targeting employees aged 50 and above for
termination.

\begin{figure}

\includegraphics{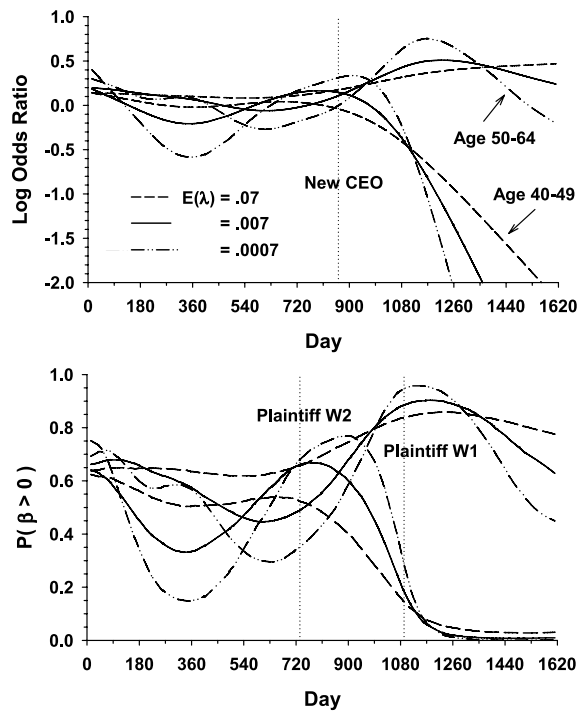}

\caption{Smooth by piecewise constant proportional
hazards model.}\label{fig:Smooth-by-piecewise}
\end{figure}

In the original case, the plaintiff's statistical expert tabulated
involuntary termination rates for each calendar quarter and each age
decade. He reported that, ``[Involuntary] separation
rates for the [period beginning at day 481] averaged a little above
three percent of the workforce per quarter for ages 20 through 49,
but jumped to six and a half percent for ages 50 through 59. The 50--59
year age group differed significantly from the 20--39 year age group
(signed-rank test, $p=0.033$, one sided).''
The plaintiff alleged and the defendant denied that the new CEO had
vowed to weed out older employees. The case was settled before trial.

In a subsequent re-analysis [\citeasnoun{KadWoo2004}], we
employed a proportional hazards model with separate, smoothly time-varying
log hazard ratios for ages 40--49, and 50--64, with ages 20--39 as the
reference category. Thus, the log hazard ratio was smooth over time
but piecewise constant over age; Figure \ref{fig:Smooth-by-piecewise}
is reproduced with permission from that paper. Our preferred model,
represented by the solid curves, had prior mean smoothness 0.007.
For this prior the posterior probability of age-discrimination
in the case of Plaintiff W1 was 0.842.

The model depicted in Figure \ref{fig:Smooth-by-piecewise} has two explanatory variables for
age, an indicator variable for age in the range 40--49 and an
indicator variable for age 50 and above (there are no employees 65 and
over in the data set). The likelihood model was proportional hazards
regression with smoothly time-varying coefficients for the two
explanatory variables. Three analyses are shown here with different
prior means for the smoothness parameter, $\lambda$. The upper panel
shows posterior means of the proportional-hazards regression
coefficients as functions of time and the smoothness parameter. As
suggested in the figure, the regression coefficients are interpretable
as instantaneous log-odds ratios with unprotected, under-40,
employees as the reference category. The second panel presents
posterior probabilities that the two regression coefficients are
positive; that is, that the termination rate is higher for the
protected subclasses compared to the unprotected class. For example, at the time of
plaintiff W2's termination, the posterior probability exceeds 80\%
that employees age 50 and above had a higher risk of termination than
the protected class.

\begin{table}[b]
\tabcolsep=0pt
\caption{Summary of three analyses of Case W}\label{tab:Wsummary}
\begin{tabular*}{\textwidth}{@{\extracolsep{\fill}}llll@{ \ }lcc@{}}
\hline
& & & & &\multicolumn{2}{c@{}}{\textbf{Plaintiff}}\\ [-7 pt]
& & & & & \multicolumn{2}{c@{}}{\hrulefill}\\
\textbf{Analysis}&\textbf{Method} &
\multirow{2}{36pt}[12pt]{\textbf{Figure of merit}}&
\multirow{2}{40 pt}[12pt]{\textbf{Treatment of age}}  &
\multirow{2}{55pt}[12pt]{\textbf{Age $\bolds\times$ time
interaction}}& \multicolumn {1}{c}{\textbf{W1}} &
\multicolumn{1}{c@{}}{\textbf{W2}}\\
\hline
\multicolumn{1}{@{}p{40 pt}}{Original expert's report} &
Frequentist & $p$-value &\multirow{2}{43 pt}{categorical: 40-up}
& none & 0.033& 0.053\\
\multirow{3}{55 pt}{\citeasnoun{KadWoo2004}} & Bayesian &
\multirow{3}{65 pt}{probability of disproportional disadvantage}
& \multirow{2}{43 pt}{categorical: 40--49, 50--64 }& smooth
& 0.84 & 0.50\\\\
& & & & \multirow{2}{58pt}{smooth/w discontinuity at day 862} &
0.88 & 0.49\\\\\\
This paper & Bayesian &
\multirow{2}{65 pt}{probability of disproportional disadvantage} &
smooth & smooth & 0.65 & 0.37\\\\\\
\multirow{3}{45pt}{Anonymous referee of this paper} &
\multirow{2}{42 pt}{Cox regression} & \multirow{2}{45pt}{$p$-value, OR, and
90\% LCL} & \multirow{2}{40 pt}{linear above 40} &
\multirow{2}{44 pt}{none but restricted to day 1000 up} &
\multicolumn{1}{l}{\multirow{3}{40 pt}{$p$:
0.041 OR: 2.04 LCL: 1.01}} & \multicolumn{1}{l@{}}{n/a} \\\\\\\\
\hline
\end{tabular*}
\end{table}

A second plaintiff, W2 aged 60 terminated on day 733, also brought
an age-discrimination suit on the theory that employees aged 60 and
above were disproportionately targeted at the time of his termination.
On that day three of eight employees (37.5\%) aged 60 and up were
terminated compared to 15 of 136 (11.0\%) employees terminated out
of all other age groups (one-sided Fisher exact test $p=0.0530$).
In our re-analysis the posterior probability of age discrimination
against employees aged 50--64 was about 50\% but did not distinguish
between employees aged 50--59 and 60--64. Our second re-analysis reported
in this paper remedies that deficiency and gives a more detailed picture
of the impact of age on the risk of discrimination; in particular,
for our preferred prior, the posterior probability of age discrimination
against 60-year old employees on day 733 is about 65\% but is only
about 37\% for 50-year old employees.

\subsection{Summary}

Table \ref{tab:Wsummary} summarizes the results of the three analyses of
case W for each of the two plaintiffs. In the first, classical,
analyses for Plaintiff W1, it is assumed that each employee in the age
groups 20--39 and 50--59 has the same chance of being involuntarily
terminated (i.e., fired) in each quarter-year after day 481. The
test of significance calculates the probability of obtaining data as
or more extreme than that observed were it true that persons in these
two age groups have the same chance of being fired in any given
quarter. The classical analysis for plaintiff W2 is somewhat
different, in that it focuses solely on what happened on the day that
W2 was fired. It conditions on both the age distribution of the
workforce at the time (eight of 144 employees 60 years old or older)
and the number fired (18) and computes the probability of three or
more of the eight older employees being fired, if employees were
equally likely to be fired.

The second analysis is based on
a model for the log odds of being fired that is continuous in time but
still assumes constancy in age categories. The analysis of this paper
relaxes this latter assumption, and allows smoothness in both age and
time. In both Bayesian analyses, the probability computed is that an
employee of a given age was more likely to be fired at a particular
time than was an employee in the unprotected 20--39 age group.

Although the classical analyses are computing probabilities in the
sample space while the Bayesian analyses are computing probabilities
in the parameter space, the stronger effect here appears to be that as
the assumptions get less rigid, there is less certainty that these
plaintiffs' cases were meritorious, as Table \ref{tab:Wsummary} shows. In view
of the tendency of Bayesian analyses to draw estimates toward each
other, this is perhaps not too surprising.

\section{Discussion}

In a nonhierarchical model, the effect of the prior can be
isolated by separately reporting the likelihood function and the
prior distribution. In particular, if the parameter space is
divided into two disjoint subsets, the likelihood ratio and the
prior odds suffice. However, in a hierarchical model such as this
one, such a separation is not possible. For this reason, we
have reported the results of changing our prior directly, in
Sections~\ref{sub:sensitivity}, \ref{sub:anisotropy} and \ref{sub:previous}.

We have presented a global analysis of involuntary terminations that
incorporates all of the data but reflects fine-grained variations
over time and age of employee. The results are somewhat sensitive
to assumptions about prior distribution of the smoothness parameter,
although not enough to materially alter the strength of evidence supporting
the plaintiff's discrimination claim in Case W. This analysis, in our view,
casts new light on the apparent patterns
in coarser-grained descriptive presentations that might be easier
for nonspecialists to grasp.

Our intent is to develop a methodology that does not require complex
assumptions about the relationship between time, age and risk of
termination. Indeed, the only structural assumption is smoothness and
the only prior opinion required has to do with the degree of
smoothness. We have suggested how that prior opinion could be elicited
by considering how rapidly the risk of termination is likely to change
over a business quarter and over a decade of age. A referee described
our analysis as ``staggeringly complex'' and ``shuddered to think what
a judge or jury would make of this approach.'' All statistical
analyses are ``staggeringly complex'' to most laypersons. We think our
responsibility as statisticians (and experts in court) is to present
our best analysis of the data, and to explain it as best as we can.

A global analysis such as this one is more powerful and more appropriate
than analyzing subsets of the data, perhaps in the form of individual
termination waves or individual business quarters, and more appropriate
than analyzing coarse aggregations such as employees aged 40 and above
compared to younger employees. The fallacy of subdividing the data
is that such analyses implicitly assume that there is no continuity
in the behavior of a firm and no difference in treatment of employees
of different ages within the same broad age category (40 and older).
We believe that the appropriate approach to possible inhomogeneities
of the age effect is to incorporate them in a global model---see,
for example, our discussion of Gastwirth's (\citeyear{Gastwirth1992}) analysis
in Valentino v. United States Postal Service [\citeasnoun{Gastwirth1992}, \citeasnoun{KadWoo2004}].

Finally, it has not escaped our notice that our analysis of Case W
has made it clear that only a subgroup of older employees, centered
around the peak at day 1275 and age 54--55, has even moderately strong
statistical evidence to support a claim of age discrimination. We
believe that this is precisely the information that the court needs
in order to determine how an award (if any) should be distributed among members
of a certified class.
\begin{supplement}[id=suppA]
\sname{Supplement A}
\stitle{Employment --- Case W}
\slink[doi]{10.1214/10-AOAS330SUPPA}
\slink[url]{http://lib.stat.cmu.edu/datasets/caseW.txt}
\sdatatype{.txt}
\sdescription{Data from two cases described in the paper {}``Hierarchical models for
employment decisions,'' by Kadane and Woodworth. A constant number of
days has been subtracted from each date to preserve
confidentiality.}
\end{supplement}

\begin{supplement}[id=suppB]
\sname{Supplement B}
\stitle{Code for calculations}
\slink[doi]{10.1214/10-AOAS330SUPPB}
\slink[url]{http://lib.stat.cmu.edu/aoas/330/supplement.zip}
\sdatatype{.zip}
\end{supplement}

\printaddresses

\end{document}